\documentclass[preprint,amsmath,amssymb,prl,epsf,rotate]{revtex4}

\usepackage{graphicx}
\usepackage{dcolumn}
\usepackage{bm}

\begin{document}

\title{A Generalized Preferential Attachment Model \\ for Complex Systems }

\author{Kazuko~Yamasaki$^{1,2}$, Kaushik Matia$^2$, Dongfeng Fu$^2$,
Sergey~V.~Buldyrev$^2$, Fabio Pammolli$^3$, Massimo Riccaboni$^3$ and
H. Eugene~Stanley$^2$ }

\affiliation{ $^1$Tokyo University of Information Sciences, Chiba City
265-8501 Japan. \\ $^2$Center for Polymer Studies and Department of
Physics, Boston University, Boston, MA 02215 USA.\\ $^3$Faculty of
Economics, University of Florence and CERM, Via Banchi di Sotto 55,
Siena 53100 Italy.  }

\date{\today}

\begin{abstract}

Complex systems can be characterized by classes of equivalency of their
elements defined according to system specific rules. We propose a
generalized preferential attachment model to describe the class size
distribution. The model postulates preferential growth of the existing
classes and the steady influx of new classes. We investigate how the
distribution depends on the initial conditions and changes from a pure
exponential form for zero influx of new classes to a power law with an
exponential cutoff form when the influx of new classes is
substantial. We apply the model to study the growth dynamics of
pharmaceutical industry.

\end{abstract}

\maketitle

\date{working paper last revised: \today}

Many diverse systems of physics, economics, and
biology~\cite{Barabasi,Sergey,City,Zipf,Satellite}, share in their
growth dynamics two basic similarities: (i) The system does not have a
steady state and is growing. (ii) Basic units are born and they
agglomerate to form classes. Classes grow in size preferentially
depending on the existing size. In the context of economic systems,
units are products, and the classes are firms. In social systems units
are human beings, and the classes are cities. In biological systems
units can be bacteria, and the classes are the bacterial colonies.

The probability distribution function $p(k)$ of the class size $k$ of
the systems mentioned above share a universal behavior $p(k) \sim
k^{-\tau}$ with $\tau \approx 2$~\cite{Barabasi,City,Zipf,Kumar}. Other
possible values of $\tau$ are discussed and reported
in~\cite{Newman}. Also, for most of the systems $p(k)$ has an
exponential cutoff which is often assumed to be a finite size effect of
the databases analyzed. Several
models~\cite{Sergey,Champernowne,Fedorowicz,Gabaix,Reed,Simon} explain
$\tau \approx 2$ but none explains the exponential cutoff of
$p(k)$. Moreover, these models describing $p(k) \sim k^{-\tau}$ are not
suitable to describe simultaneously systems for which $p(k) \sim
\exp(-\gamma k)$. Here we present a model with simple set of rules to
describe $p(k)$ for the entire range of $k$, i.e., power law with an
exponential cutoff. We show that the exponential cutoff of the power law
is not due to finite size but an effect of the initial conditions from
which the system starts to evolve. We also show that the functional form
of $p(k)$ depends on the initial conditions of our model and changes
from a pure exponential to a pure power law (with $\tau = 2$) via a
power law with an exponential cutoff. We justify our model by empirical
analysis of a recently constructed pharmaceutical industry database
(PHID)~\cite{Pammolli,Matia}.

We now present a model, which has the following rules:
\begin{enumerate}
\item At time $t=0$ there exists $N$ classes, each with a single unit.
\item At each simulation step:
\begin{itemize}
\item (a) With probability $b \ ( 0 \le b \le 1 )$ a class with a single
unit is born.
\item (b) With probability $\lambda \ ( 0 < \lambda \le 1 )$ a randomly
selected class grows one unit in size. The selection of the class that
grows is done with probability proportional to the number of units it
already has [``preferential attachment''].
\item (c) With probability $\mu \ ( 0 < \mu \le 1, \mu < \lambda )$ a
randomly selected class shrinks one unit in size. The selection of the
class that shrinks is done with probability proportional to the number
of units it already has [``preferential detachment''].
\end{itemize}
\end{enumerate}
In the continuum limit the proposed growth mechanism gives rise to a
master equation of $p(k,t_i ,t)$ which is the probability, for a class
$i$ born at simulation step $t_i$, to have $k$ units at step $t$:.
\begin{gather}
\begin{split} 
 & \frac{{\partial p(k,t_i ,t)}}{{\partial t}} = \lambda \frac{{(k -1)}}{{g(t)}}p(k - 1,t_i ,t)\\ 
 &   + \mu \frac{{(k + 1)}}{{g(t)}}p(k + 1,t_i ,t) 
    - (\lambda  + \mu )\frac{k}{{g(t)}}p(k,t_i ,t)  
\label{eqmaster}
\end{split}
\end{gather}
where $g(t) \equiv N + (\lambda -\mu+b)t$ is the total number of units
at simulation step $t$ and $p(1,t_i ,t_i ) =
1$. Equation~(\ref{eqmaster}) is the generalization of the master
equation of birth and death processes~\cite{Reed}. The analytical
solution of Eq.~(\ref{eqmaster}) is given by
\begin{equation}
p(k,t) = \frac{N}{{N + b t}}p(k,0,t), {\kern 10pt} + \frac{{b }}{{N + b t}}\int_0^t {dt_i {\kern 1pt} } p(k,t_i ,t)
\label{eqsol1}
\end{equation}
where the functional form of $p(k,t_i,t)$ is given in~\cite{eqref}. The
lengthy derivation of the full solution of eq.~\ref{eqmaster} which is a
power law(the second term of eq.~\ref{eqsol1}) with an exponential
cutoff (the first term of eq.~\ref{eqsol1}) will be presented elsewhere,
here we present simulation results.

\begin{figure*}
\narrowtext
\begin{center}
\includegraphics[scale=0.38,angle=-90]{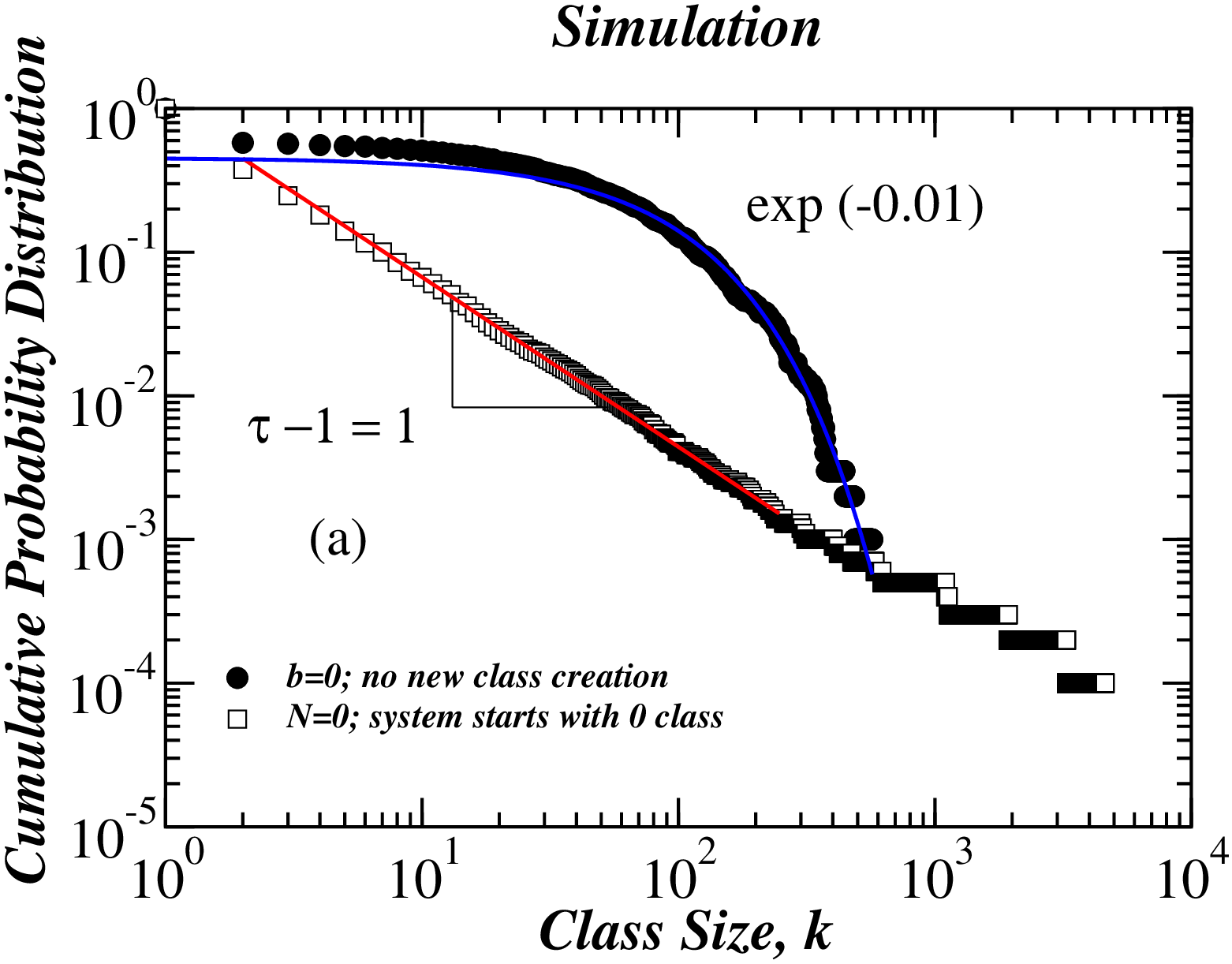}
\includegraphics[scale=0.38,angle=-90]{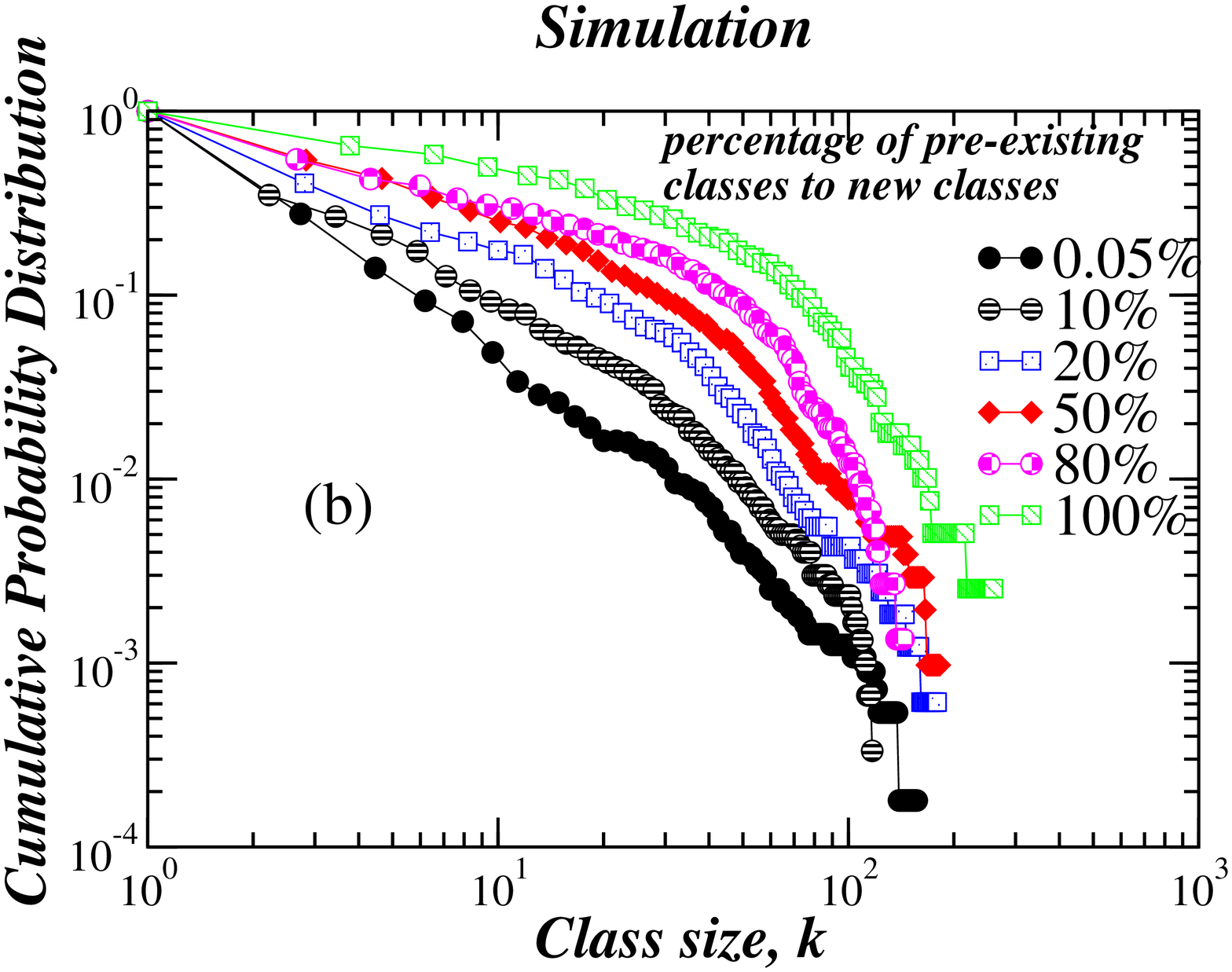}
\end{center}
\caption{ Simulation results of the model. (a) Symbols are data points
from simulation, solid lines are regression fits. We observe for $b=0$
(i.e. no class creation ) cumulative probability distribution is a pure
exponential while for $N=0$ (i.e. we start with zero initial class ) a
pure power law $k^{-(\tau -1)}$ with exponent $\tau = 2$. (b) We observe
that as we change the ratio of number of pre-existing classes to the new
born classes $p(k,t)$ changes from a pure power law to a pure
exponential. }
\label{simulation}
\end{figure*}

First we discuss two limiting solutions of Eq.~(\ref{eqmaster}).
\begin{itemize}
\item {\it Case i} : No new classes are born ($b=0$). The growth of the
system is solely due to the preferential attachment of new units to the
pre-existing $N$ classes. In this case
(Fig~\ref{simulation}a)~\cite{Reed}
\begin{equation}
p(k) \sim e^{-\frac{kN}{t(\lambda -\mu)}}.
\end{equation} 
This limiting case can be considered as one of the initial condition of
the model where birth or death of classes are {\it not} allowed. We observe that
this initial condition results in a pure exponential distribution of the
number of units inside classes.
\item {\it Case ii} : At $t=0$, $N=0$, and new classes are born with
probability $b \neq 0$. In this case, for large times $p(k)$ is a pure
power law
\begin{equation}
p(k) \sim k^{-\tau}, \quad \tau = 2. 
\end{equation}

This limiting case can be considered as another different initial
condition of the model where birth or death of classes are allowed
starting from $N=0$ classes. We observe that this initial condition
results in a pure power law distribution of the number of units inside
classes.

This case is identical to the Simon model~\cite{Simon} and can be
understood by the following arguments. From case (i) we know that when
the number of classes remains constant, $p(k)$ decays exponentially with
$k$. The power law of case (ii) is the effect of superposition of many
exponentials with different decay constants, each resulting from classes
born at different times (Fig~\ref{simulation}a).

\end{itemize}

We next present a mean field interpretation of the result $\tau\approx
2$. At any moment $t_0$ the number of units in the already-existing
classes is $g(t_0)$. Suppose a new class consisting of one unit is
created at time $t_0$. According to rules 2b, 2c, the growth rate is
proportional to $1/g(t_0)$. Neglecting the effect of the influx of new
classes on $g(t_0)$, the average size $k$ of this class born at $t_0$ is
proportional to $1/g(t_0)$. So the classes which were born at times $t >
t_0$ remain smaller than the classes born earlier. If we sort the
classes according to their size, the rank $R(k)$ of a class is
proportional to the time of its creation $R(k) \propto t_0$. Thus $k
\sim 1/g(t_0) \sim 1/t_0 \sim 1/R(t_0)$ and we arrive to the standard
formulation of the Zipf's law~\cite{Zipf} according to which the size of
a class $k$ is inversely proportional to its rank. If we take into
account the decrease of the growth rate with the influx of new classes,
one can show after some algebra $k\sim R^{-(\lambda -\mu)/(\lambda -
\mu+b)}$, which includes $k\sim R^{-1}$ as a limiting case for $b
\rightarrow 0$. Since $R(k)$ is the number of classes whose size is
larger than $k$, we can write in the continuum limit $R(k) \sim
\int_k^{\infty} p(k) dk $ and hence $p(k) \sim k^{-2-b/(\lambda -
\mu)}$.

The full solution of Eq.~(\ref{eqmaster}), a power law with an
exponential cutoff, can be interpreted using the following arguments. We
start with $N$ classes which are colored red, and let the newly born
classes be colored blue. Due to the preferential attachment rule, the
red classes remain on average larger than the blue classes. Thus for
large $k$, $p(k)$ is governed by the exponential distribution of the red
classes ({\it Case i}) while for small $k$, $p(k)$ is governed by the
power law distribution of the blue classes ({\it Case ii})
(Fig.~\ref{simulation}b).

\begin{figure*}
\narrowtext
\begin{center}
\includegraphics[scale=0.48,angle=-90]{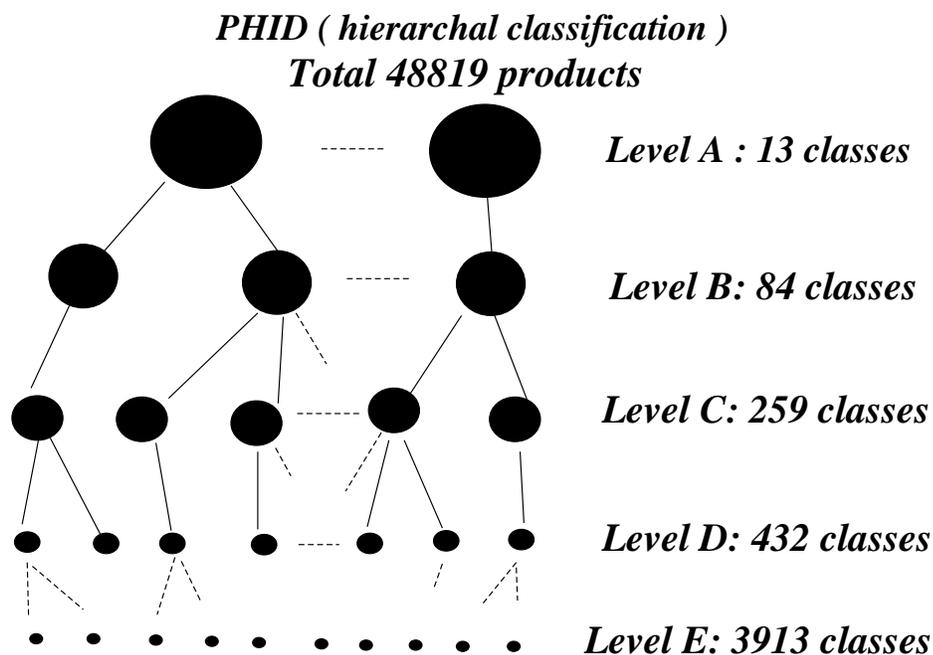}
\end{center}
\caption{In the pharmaceutical industry, products can be classified
according to five levels. When a particular product arrives in the
market, it is labeled under any one of the 13 classes of the level A, 84
classes of level B, and so on. Since the 19th century, the number of
classes of level A or B has remained constant even though the number of
products within each class had a dramatic growth. Over the period of our
empirical analysis the number of classes in levels C and D increased by
3\% and 5\% respectively. Products can also be grouped into firms which
markets them (classification level E). In the figure we give the number
of classes in each level in 1991.}
\label{figphid}
\end{figure*}

Now we apply this model to describe the statistical properties of growth
dynamics of business firms in pharmaceutical industry. PHID records
quarterly sales figures of 48 819 pharmaceutical products commercialized
in the European Union and North America from September 1991 to June
2001. The products in PHID can be classified in five different
hierarchal levels A, B, C, D, and E
(Fig.~\ref{figphid})~\cite{note1}. Each level has a different number of
classes, and different initial conditions (Table~\ref{ncluster}).

We observe that there are positive correlations between the number of
units (products) appearing or disappearing per year and the number of
units in the classes at a particular hierarchal level
(Table~\ref{corr}). This empirical observation supports preferential
birth or death mechanism (rules 2b, 2c) used in our model.

\begin{figure*}
\narrowtext
\begin{center}
\includegraphics[scale=0.68,angle=-90]{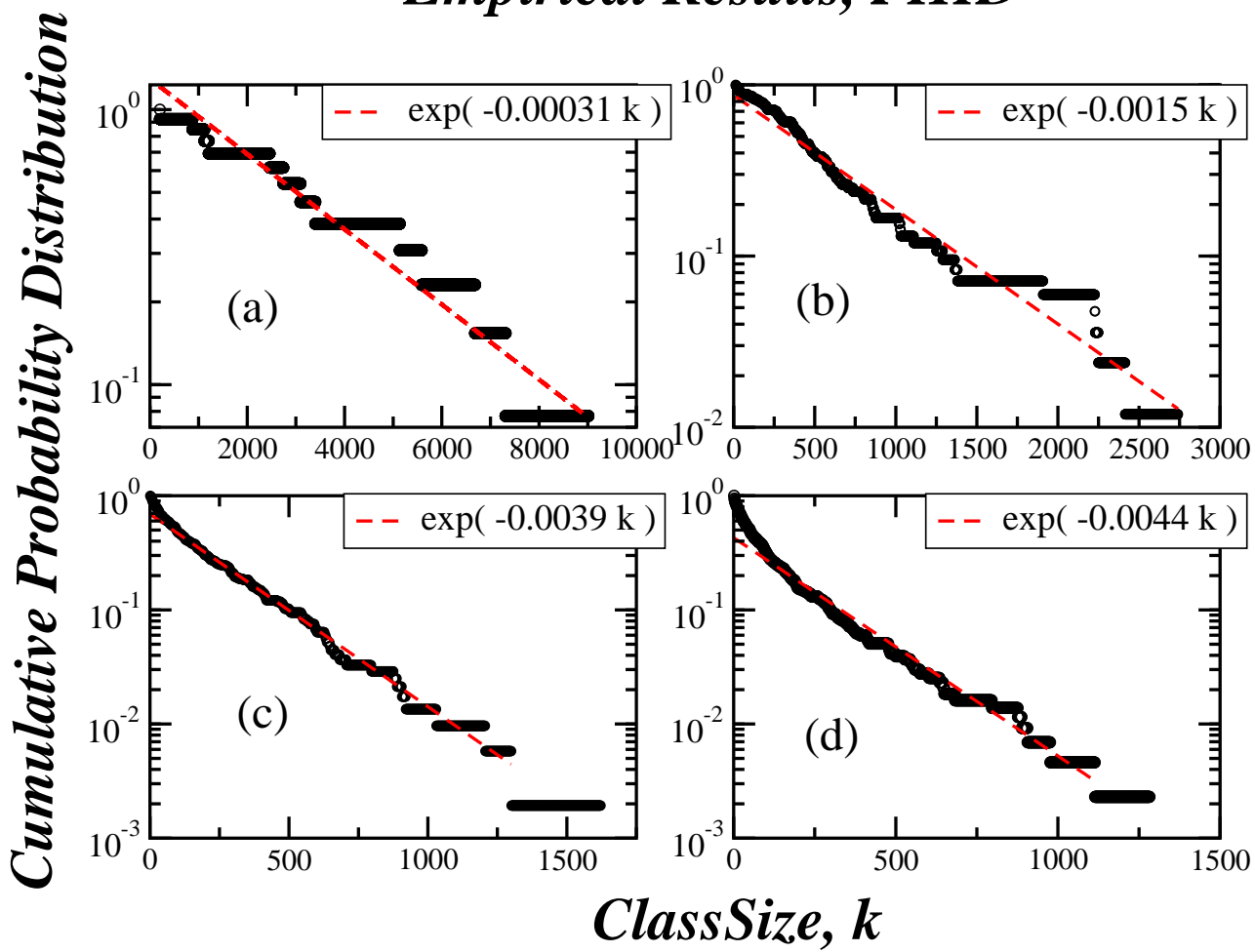}
\end{center}
\caption{Figures (a)$\sim$(d) corresponds to levels (A)$\sim$(D)
respectively. Products in the pharmaceutical industry are classified
into levels A, B, C and D. Levels A and B have fixed numbers of classes,
the number of classes in levels C and D increases by 3\% and 5\%
respectively over the period of our analysis. For instance, for level A
(fig. 3 a) which contains only 13 classes, the distribution is estimated
from 13 random interger numbers which corresponds to classifying 48,819
products in 13 classes. Symbols represent data points in each level
(a)$\sim$(d) while solid lines are predictions of the model. Cumulative
probability distributions for all levels are pure exponentials as
predicted by the model. }
\label{exp}
\end{figure*}

For levels A and B where the number of classes did not change we obtain
an exponential distribution (Figs.~\ref{exp}a,~\ref{exp}b) as predicted
by limiting {\it Case i} of the model. For levels C and D a weak
departure from the exponential functional form
[Figs.~\ref{exp}c,~\ref{exp}d] is due to the slight growth in the number
of classes.

\begin{figure*}
\narrowtext
\begin{center}
\includegraphics[scale=0.38,angle=-90]{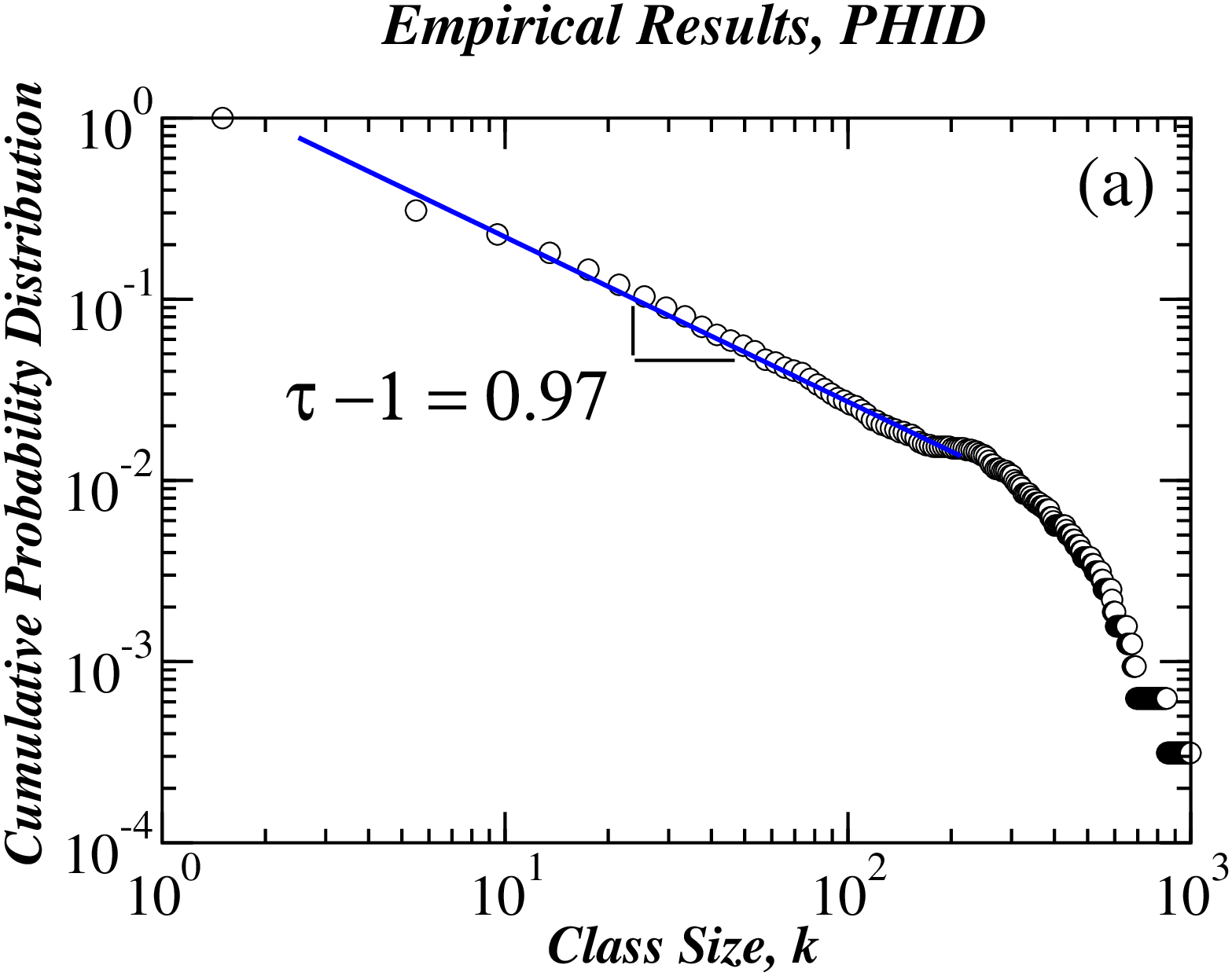}
\includegraphics[scale=0.38,angle=-90]{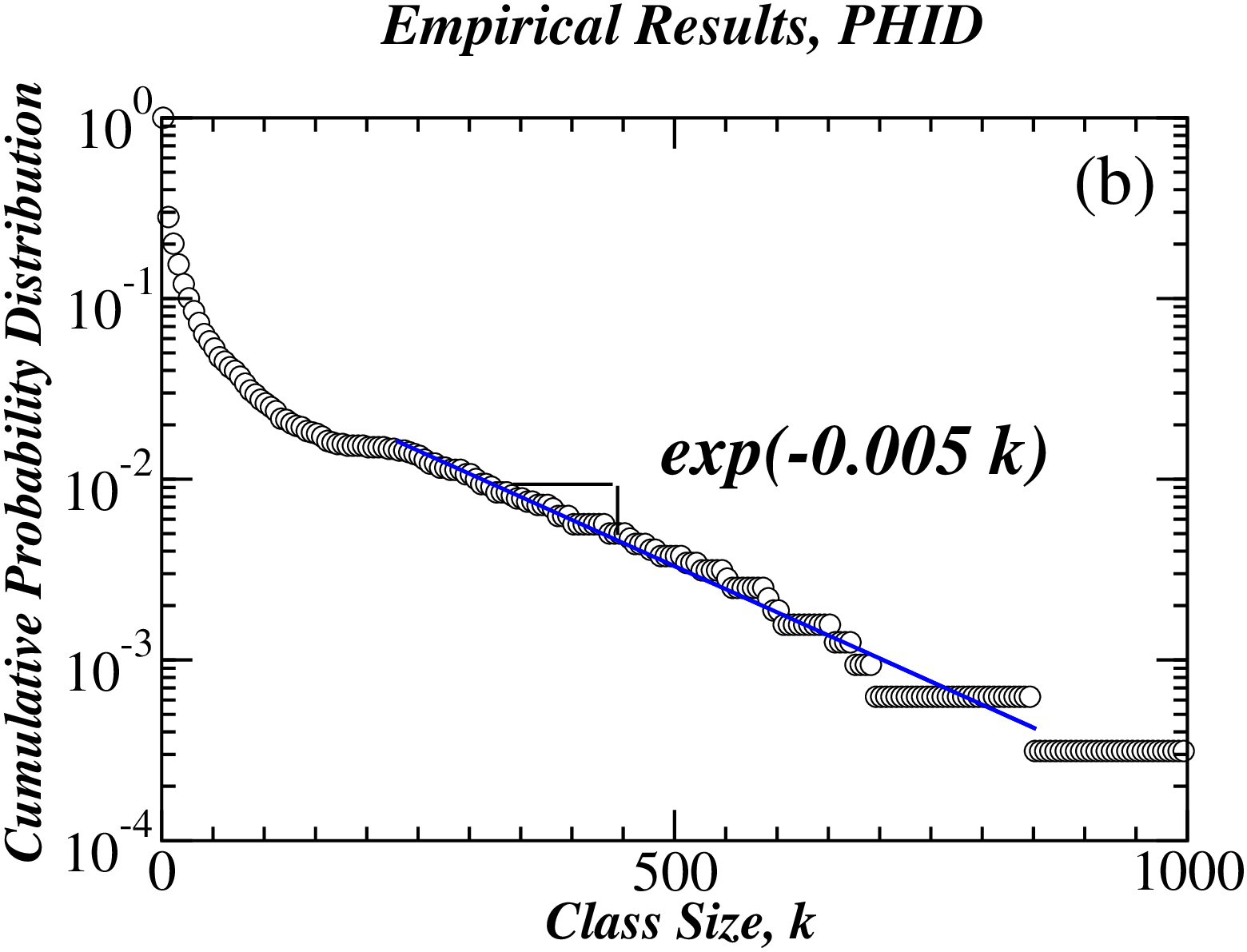}
\end{center}
\caption{Empirical results from PHID level E. The classes analyzed here
are the firms. Circles are data points, solid lines are regression
fits. (a) Log-log plot of cumulative probability distribution of the
class sizes show a power law decay $k^{-(\tau -1)}$ with $\tau \approx 2$
for $k<200$. (b) Log-linear plot of cumulative probability distribution
show the exponential decay for $k>200$.}
\label{power}
\end{figure*}

The full solution predicted by our model, i.e., the initial power law
followed by the exponential decay of $p(k)$ is observed empirically for
level E (Fig.~\ref{power}). For level E we observe a power law with
$\tau =1.97$ for $k<200$, and an exponential cutoff for $k>200$. From
the discussion above with red and blue classes we may infer that the
exponential part of $p(k)$ arises from pre-existing firms, while the
power law part of $p(k)$ represents the young firms that enter the
market. We conclude by noting that our model is in agreement with
empirical observation where we observe $p(k)$ to be pure exponential or
a power law with an exponential cutoff. Our analysis also sheds light on
the emergence of the exponent $\tau \approx 2$ observed in certain
biological, social and economic systems.

\newpage

\begin{table}

\begin{tabular}{|c|c|c|c|c|c|}
\hline
Level & A & B & C & D & E \\
\hline 
total number of   &13 & 84 & 259 & 432 & 3913  \\
classes in each levels & & & & &  \\
\hline
number of classes    & 0 & 0  & 8   &  20 &  458 \\ 
born in each level    & & & & &  \\
\hline
number of classes  & 0 & 0  & 0   &   0 &  252 \\ 
died in each level & & & & & \\
\hline
\end{tabular}  
\caption{Two different initial conditions for classes in PHID levels:
(i) For levels A and B we have no birth or death of classes. System
grows with the birth or death of units to pre-existing $N$ classes (13
for level A and 84 for level B). (ii) For levels C and D system grows
not only with the birth or death of classes but also with birth and
death of units inside classes. }
\label{ncluster}
\end{table}

\begin{table}

\begin{tabular}{|c|c|c|c|c|c|}
\hline
Level & A & B & C & D & E  \\
\hline 
correlation between number   & 0.93 & 0.87 & 0.84 & 0.82 & 0.70\\ 
of units born and existing  & & & & & \\
number of units in classes & & & & & \\
\hline
correlation between number  & 0.88 & 0.86 & 0.80 & 0.78 & 0.75\\ 
of units died and existing  & & & & & \\
number of units in classes & & & & & \\
\hline 
\end{tabular} 
\caption{Correlation of birth and death of units with existing number of
units in classes for each level in PHID. This observed correlation
justifies the preferential birth or death of units which is rule 2 b and
2 c of our model.}
\label{corr}
\end{table}

\newpage


\eject

\eject

\eject

\eject

\newpage



\begin{thebibliography}{1}
\bibitem{note1} The different levels A, B, C, D of PHID are the four
different anatomical therapeutic chemical (ATC) levels. Classes in ATC 1
are organs of the body, classes in ATC 2 are therapeutic preparations,
classes in ATC 3 and 4 are the chemicals and compounds respectively used
in preparing the products. For level E the classes are the firms like
Merck, Glaxo etc.
\end{thebibliography}
\end{document}